\documentstyle[psfig]{l-aa}

\begin{document}

   \thesaurus{06         
              (03.11.1;  
               16.06.1;  
               19.06.1;  
               19.37.1;  
               19.53.1;  
               19.63.1)} 
   \title{Star clusterings in the Carina complex:\\ $UBVRI$
 photometry of NGC~3324 and Loden~165
   \thanks{Based on observations taken at ESO La Silla.}}

   \author{G. Carraro\inst{1,2}, F. Patat\inst{2},   
          and H. Baumgardt\inst{3} 
		  }

   \offprints{G. Carraro ({\tt carraro@pd.astro.it})}

   \institute{Dipartimento di Astronomia, Universit\'a di Padova,
	vicolo dell'Osservatorio 5, I-35122, Padova, Italy
        \and
	European Southern Observatory, 
 Karl-Schwartzschild-Str 2, D-85748 Garching b. M\"unchen, Germany
   		\and
        Department of Mathematics and Statistics, University of
	Edimburgh, Edinburgh EH9 3JZ,	
	UK\\
        e-mail: {\tt 
carraro@pd.astro.it,fpapat@eso.org,holger@maths.ed.ac.uk}
             }

   \date{Received ; accepted}

   \maketitle

   \markboth{Carraro et al}{Observations of NGC~3324 and Loden~165}

   \begin{abstract}
We report on $UBVRI$ photometry of two $5^{\prime} \times
5^{\prime}$ fields in the region of the young open cluster NGC~3324. 
One of our fields covers the 
core region, while the other is closer to the tidal radius of the cluster.
Our study provides the first CCD photometry of NGC~3324.
We find that the cluster is very young and probably
contains several pre Main Sequence
(MS) stars. 25 members are identified on the basis of their position
in the $(U-B)$ vs $(B-V)$ diagram.
We investigate the relation of the red super-giant HD~92207 with NGC 3324,
suggesting that it probably does not belong to the cluster.\\
Our second field is close to Loden 165, a possible cluster of stars
that has never been studied
so far. We show that this object is a probable open cluster, much older than
NGC~3324 and much closer to the Sun.

   \end{abstract}

%

\section{Introduction}
Aiming at collecting a homogeneous and complete sample of all the star
clusters in the Carina complex (Janes et al 1988), we have carried out
a program dedicated at obtaining 
high quality multicolor CCD photometry of all the star
clusterings known or presumed to lie close to $\eta$ Carinae.\\
We already reported on Bochum~9, 10 and 11 in Patat \& Carraro (2001).
In this work we concentrate on the results we have obtained
for NGC~3324 and Loden~165, while the remaining clusters
(see Feinstein 1995 for a review on the subject) 
will be discussed in Romaniello et al (2001).\\
NGC~3324 is particularly interesting
because it is believed to be extremely young and to have a significant pre-MS
population. 
Moreover the membership of the red super-giant star
HD~92207 (HIP 52004) to NGC~3324 is still uncertain. 
Although the projected distance of HD~92207
from the cluster center seems much larger than the cluster radius, there
are reasons to believe that this star is a member of NGC~3324,
although we are far from having a firm answer to this problem
(Baumgardt et al.\ 2000).\\
As for Loden~165 (C1035-584 according to IAU), 
no data are available in the literaure to
our knowledge. Therefore it is not clear whether it is a 
star cluster or not, and whether it is related 
or not to the Carina complex.\\
The layout of the paper is as follows: Section~2 presents
the data acquisition and reduction. In Section~3 we discuss the open
cluster NGC~3324 and the relation of HD~92007 to it.
Section~4 is dedicated to Loden~165, a triplet
of stars that Lod\`en (1973) suggested to be the signature
of a possible star cluster, demanding further investigation.\\  
Our conclusions
are summarized in Section~5.

\begin{table}
\caption{Basic parameters of the observed objects.
Coordinates are for J2000.0 equinox. For Loden~165, the star  
HD~303088 is used as center of the system.}
\begin{tabular}{ccccc}
\hline
\hline
\multicolumn{1}{c}{Name} &
\multicolumn{1}{c}{$\alpha$}  &
\multicolumn{1}{c}{$\delta$}  &
\multicolumn{1}{c}{$l$} &
\multicolumn{1}{c}{$b$} \\
\hline
& $hh:mm:ss$ & $^{o}$~:~$^{\prime}$~:~$^{\prime\prime}$ & $^{o}$ & $^{o}$ \\
\hline
NGC~3324       & 10:37:18.8 & -58:37:36.2 & 286.22 & -0.18\\
Loden~165      & 10:35:55.8 & -58:44:02.7 & 286.12 & -0.36\\
\hline\hline
\end{tabular}
\end{table}

\section{Observations and Data Reduction}
Observations were conducted at La Silla on April 13-16, 1996, using the 
imaging Camera ( equipped with a TK coated 512 $\times$ 512 pixels CCD \#33) 
mounted at the Cassegrain focus of the 0.92m ESO--Dutch telescope. All 
nights, with the only exception of the first half of April 14, were 
photometric with a seeing ranging from 1$^{\prime\prime}$ to 
2$^{\prime\prime}$. 
The scale on the chip is 0$^{\prime\prime}.44$ and the array covers
about 3$^\prime$.3 $\times$ 3$^\prime$.3 on the sky. Due to the projected
diameter of the objects and the relatively small field of view,  
it was necessary to observe four fields for the same object.
Additional details on the observations are given in Table~2, 
while the  covered fields are shown in Fig.~1.

\begin{table}
\tabcolsep 0.10truecm
\caption{Journal of observations of NGC~3324 (April 16, 1996) and
Loden 165 (April 15, 1996).}
\begin{tabular}{ccccccc} \hline
 & \multicolumn{3}{c}{NGC~3324} & \multicolumn{3}{c}{Loden 165} \\
Field & Filter & Exp. Time & Seeing         & Filter & Exp. Time & seeing \\
      &        & (sec)     & ($\prime\prime$) &        &  (sec)    & ($\prime\prime$)\\
      &        &           &                  &        &           & \\  
   $\#1$ & U &   10 &  1.2 & I &   20 &  1.8\\
	 & B &    3 &  1.1 & I &  120 &  1.8\\
	 & B &  120 &  1.1 & R &    5 &  1.8\\
	 & V &    3 &  1.1 & R &   60 &  1.8\\
	 & V &   60 &  1.1 & V &   10 &  1.8\\
	 & R &    3 &  1.1 & V &  120 &  1.8\\
	 & R &   60 &  1.1 & B &   15 &  1.8\\
	 & I &    3 &  1.1 & B &  300 &  1.8\\
	 & I &  120 &  1.1 & U &  600 &  1.9\\
 $\#2$ & I &    3 &    1.1 & U &  600 &  2.0\\
	 & I &  120 &  1.1 & B &   15 &  1.8\\
	 & R &    3 &  1.1 & B &  300 &  1.8\\
	 & R &   30 &  1.1 & V &   10 &  1.7\\
	 & V &    3 &  1.1 & V &  120 &  1.7\\
	 & V &   60 &  1.1 & R &    5 &  1.5\\
	 & B &    3 &  1.1 & R &   60 &  1.6\\
	 & B &  120 &  1.1 & I &   20 &  1.6\\
	 & U &   10 &  1.1 & I &  120 &  1.7\\
 $\#3$ & U &   10 &    1.2 & I &   20 &  1.5\\
	 & B &    3 &  1.2 & I &  120 &  1.5\\
	 & B &  120 &  1.2 & R &    5 &  1.5\\
	 & V &    3 &  1.2 & R &   60 &  1.6\\
	 & V &   30 &  1.2 & V &   10 &  1.8\\
	 & R &    3 &  1.3 & V &  120 &  2.0\\
	 & R &   30 &  1.3 & B &   15 &  1.8\\
	 & I &    3 &  1.3 & B &  300 &  2.3\\
	 & I &  120 &  1.3 & U &  600 &  1.9\\
 $\#4$ & I &    3 &    1.5 & U &  600 &  1.9\\
	 & I &  120 &  1.4 & B &   15 &  1.6\\
	 & R &    3 &  1.5 & B &  300 &  1.6\\
	 & R &   30 &  1.4 & V &   10 &  1.7\\
	 & V &   30 &  1.4 & V &  120 &  1.7\\
	 & V &   60 &  1.3 & R &    5 &  1.7\\
	 & B &    3 &  1.4 & R &   60 &  1.7\\
	 & B &  120 &  1.4 & I &   20 &  1.7\\
	 & U &   10 &  1.5 & I &  120 &  1.5\\
\hline
\end{tabular}
\end{table}

To allow for a proper photometric calibration and to asses the night quality, 
the standard fields RU149, PG1323, PG1657, SA~109 and SA~110 (Landolt 1992)
were monitored each night. Finally, a series of flat--field frames on the
twilight sky were taken. The scientific exposures have been flat--field and
bias corrected by means of standard routines within {\it IRAF}. Further 
reductions were performed using the DAOPHOT package (Stetson 1991) under the
{\it IRAF} environment.

The instrumental magnitudes have been transformed into standard Bessel UBVRI
magnitudes using fitting coefficients derived from observations of the
standard field stars from Landolt (1992), after including exposure time
normalization and airmass correction. Aperture corrections have
also been applied. Typical RMS errors in the zero points amount at 0.02 mag.
More details on the data
reduction have been presented in Patat \& Carraro (2001).

\begin{figure*}
\centerline{\psfig{file=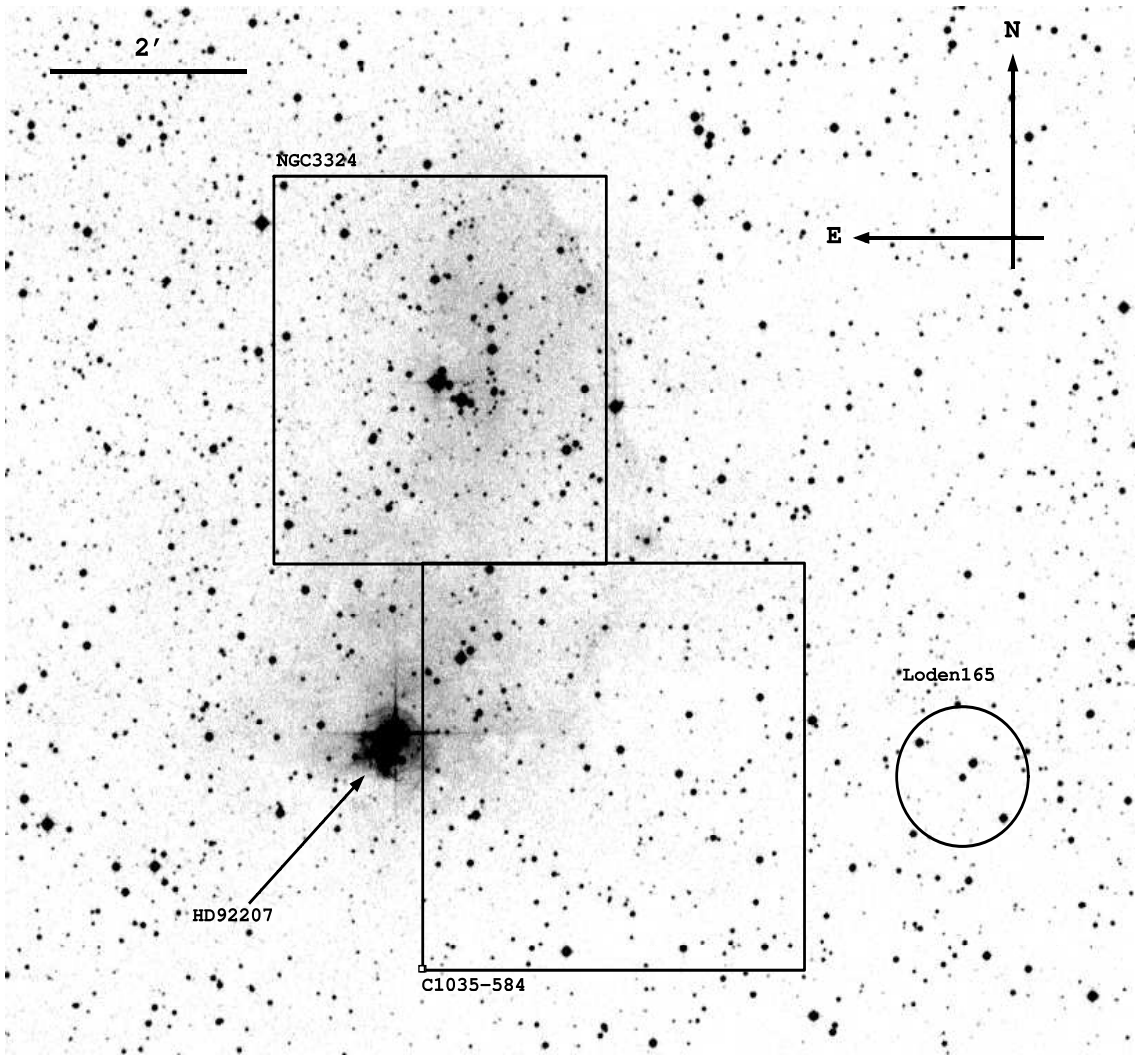,width=15cm,height=14cm}}
\caption{DSS map of the  region around NGC 3324. The region of the
center is covered by
by the upper left square, which is a mosaic of the observed fields,
whereas the lower
right square defines the 
mosaic of the fields covered by our photometry for the outer cluster parts.
The circle
confines the three stars observed by Lod\'en (1975). The bright
star on the left of our second field is HD~92207, which is presumed to be part of
NGC~3324. Note
the strong nebulosity which encloses the cluster and HD~92207.}
\end{figure*}

\section{The open cluster NGC~3324}
NGC~3324 (OCL-819) is a compact star clusters embedded
in a filamentary elliptically shaped nebulosity which
forms a bridge connecting the cluster core with HD~92207
(see Fig.~1). The complex is believed to be located inside
the Carina spiral arm.\\

\subsection{Previous results}
NGC~3324 was firstly studied by Moffat \& Vogt (1975),
who provided photoelectric photometry for 12 stars 
within $1^{\prime}$ of the cluster core. 
The authors note that the cluster is surrounded
by a large nebulosity (the HII region G~31). The strong HII 
emission supports the idea that the cluster is rather young,
together with the possible presence of contracting pre MS stars.
According to Moffat \& Vogt (1975) the cluster is located 3.28
kpc from the Sun, and suffers from a mean reddening E$(B-V)$~=~0.45$\pm$0.05.\\
Later this cluster was investigated in detail 
by Clari\'a (1977), who  obtained $UBV$ photo-electric photometry for 45 stars
in a region of about  $6^{\prime} \times 6^{\prime}$,
extending the photometry down to V=14. From  this study,
Clari\'a
concluded that NGC~3324 is 
very young ($2.2 \times 10^{6}$ yrs), has a mean color excess of  
E$(B-V)$~=~0.47$\pm$0.08, and is 3.1~kpc away from the Sun,
basically confirming  Moffat \& Vogt (1975) results.\\
Moreover Clari\'a discusses the membership of HD~92207 to
NGC~3342. He suggests that the star is a foreground red supergiant
($V$~=~5.46, $(B-V)$~=~0.50, $(U-B)$~=~$-$0.24) ,
since it is far away from the cluster and appears to have a somewhat
lower distance modulus.
On the other hand Forte (1976) comes to a different conclusion, emphasizing
the possible membership since the star seems to be part of the same
complex of NGC~3324, at least looking at the strong gaseous structure
surrounding both the cluster and the star.
The same conclusion is drawn by Baumgardt et al.\ (2000).
In fact the cluster has an angular diameter of
$5^{\prime}$ (Lynga 1987),
which means a core radius of about $3^{\prime}$ and a tidal radius
of about $12^{\prime}$.  
Since HD~92207 (HIP~52004) lies about $7^{\prime}$
southwards of NGC~3324, it stays  inside the cluster tidal
radius.
This might lead to the suggestion that  HIP~52004 is a probable
member of NGC~3324. 
This cluster has in fact a very concentrated core, and the crossing time in
the center should be very short. From time to time some stars might probably 
be kicked out of the cluster, and this can even happen to the most massive star.\\

\begin{figure}
\centerline{\psfig{file=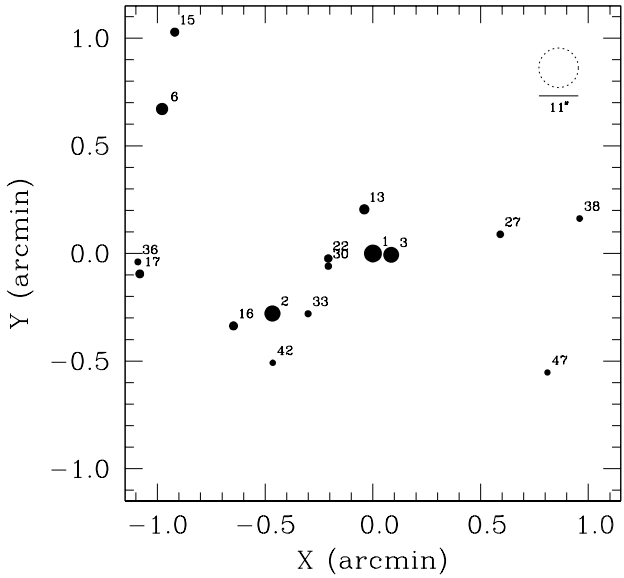,width=9cm,height=9cm}}
\caption{Map of all stars brighter than V~=~14.5 in NGC~3324. 
Note that all the star groups usually blended
in previous photo-electric photometries, are now resolved. The size of the circles
is proportional to the magnitude of the stars. The upper right dashed
circle is the size of the diaphragm ( $11^{\prime\prime}$) in the photo-electric photometry
of Moffat \& Vogt (1975).}
\end{figure}

\subsection{The present study}
We provide $UBVRI$ photometry for  1002 stars in the region of
NGC~3324, up to about V~=~20.  Our photometry is 
basically consistent
with previous ones for the stars in common (12 in total). In detail we find:

\[
V_{CPB} - V_{MV} = 0.028\pm0.095 
\]

\[
(B-V)_{CPB} - (B-V)_{MV} = 0.056\pm0.078 
\]

\[
(U-B)_{CPB} -(U-B)_{MV} = 0.080\pm 0.107
\]

\noindent 
We emphasize that we did not consider Moffat \& Vogt (1975)
stars $\#6$ and $\#9$, since they are blended. 
As for Clari\'a (1977), the comparison of the common stars
(see Table~3) provides:

\[
V_{CPB} -V_{C} = 0.003\pm0.045 
\]

\[
(B-V)_{CPB} -(B-V)_{C} = 0.021\pm0.047 
\]

\[
(U-B)_{CPB} -(U-B)_{C} = 0.082\pm0.095 
\]

\noindent
From this comparison we exclude stars $\#$ 41, 44 and 45,
which deviate more significantly (see Table~3), probably because they are either at
the limit of the Clari\'a (1977) photometry or blended (see Fig.~2).
In all the above equations 
$CPB$ stands for the present study, $MV$ for Moffat \& Vogt (1975)
and $C$ for Clari\'a (1977).\\

In both cases the differences are not negligible.
In fact Moffat \& Vogt (1975) adopted a $11^{\prime\prime}$ diaphragm 
(Moffat, private communication), while Clari\'a (1977) a $14^{\prime\prime}$ diaphragm
(Clari\'a, private communication).
Since the cluster is rather compact, and several stars have mean separation  less
than $5^{\prime\prime}$, we expect that stars blending is effective and
is probably the cause of the reported differences. We cannot however exclude
possible deviations due to differences in the adopted photometric systems.

\begin{table*}
\tabcolsep 0.5cm
\caption{Photometry of the stars identified as probable members in the
field of NGC~3324. C indicates Clari\'a (1977) numbering.}
\begin{tabular}{ccccccccc}
\hline
 $ID$  &  $C$ &   $X(pixel)$& $Y(pixel)$ &  $V$ &   $(B-V)$ &  $(U-B)$  &  $ (V-R)$ &   $(V-I)$ \\
\hline
 1  &  2  &   462.14&    37.15&    8.236&    0.129&   -0.833&    0.128&    0.380 \\
 2  &  4  &   398.57&    -0.87&    9.018&    0.205&   -0.772&    0.160&    0.319 \\
 6  &  9  &   328.74&   128.70&   10.838&    0.258&   -0.670&    0.223&    0.425 \\
 7  & 11  &   494.64&   283.34&   11.006&    0.198&   -0.682&    0.203&    0.377 \\
 13 & 19  &   456.71&    65.10&   11.789&    0.260&   -0.610&    0.216&    0.442 \\
 15 & 31  &   336.71&   177.33&   12.371&    0.286&   -0.409&    0.197&    0.410 \\
 16 & 33  &   373.93&    -8.73&   12.461&    0.219&   -0.516&    0.189&    0.365 \\
 17 & 35  &   314.59&    24.20&   12.587&    0.394&   -0.238&    0.325&    0.642 \\
 20 & 36  &   110.61&   291.96&   12.773&    0.244&   -0.361&    0.168&    0.350 \\
 22 &     &   433.90&    33.98&   12.838&    0.315&   -0.171&    0.233&    0.472 \\
 26 &     &   179.05&  -235.51&   13.264&    0.331&   -0.230&    0.249&    0.534 \\
 29 &     &   100.02&   289.55&   13.530&    0.372&    0.149&    0.425&    0.760 \\
 30 &     &   433.95&    29.10&   13.459&    0.266&   -0.268&    0.070&    0.456 \\
 31 & 41  &   406.10&   344.81&   13.547&    0.370&   -0.268&    0.268&    0.531 \\
 32 &     &   766.83&   461.47&   13.493&    0.466&    0.178&    0.302&    0.528 \\
 33 & 44  &   421.13&    -1.04&   13.696&    0.364&   -0.145&    0.274&    0.556 \\
 37 &     &   460.13&   294.91&   13.974&    0.578&    0.418&    0.396&    0.761 \\
 38 &     &   593.03&    59.27&   13.955&    0.521&    0.164&    0.315&    0.597 \\
 39 &     &   645.97&   327.91&   14.005&    0.527&    0.288&    0.315&    0.571 \\
 40 &     &   571.66&   234.97&   14.117&    0.611&    0.359&    0.473&    0.880 \\
 41 &     &   531.51&   210.38&   14.134&    0.528&    0.206&    0.392&    0.753 \\
 42 & 45  &   398.77&   -32.10&   14.148&    0.383&    0.204&    0.268&    0.526 \\
 43 &     &    16.40&  -162.82&   14.172&    0.485&    0.210&    0.295&    0.575 \\
 51 &     &   460.15&   402.78&   14.518&    0.627&    0.472&    0.370&    0.639 \\
 61 &     &   300.85&   208.82&   14.891&    0.462&    0.331&    0.315&    0.634 \\
\hline
\end{tabular}
\end{table*}

CCD photometry allowed us to resolve blended stars and to obtain fairly precise
measurements. In fact
we find that the photometric errors amount at 0.03, 0.04, 0.04, 0.0.4 
and 0.05, at V~=~ 12, 14, 16, 18 and 20, respectively.

\begin{figure}
\centerline{\psfig{file=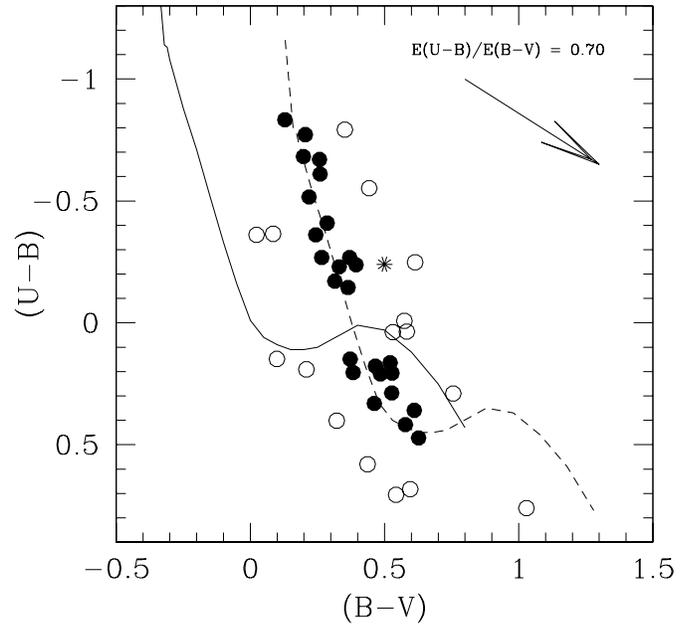,width=9cm,height=9cm}}
\caption{Two colors diagram  of NGC~3324. The arrow indicates
the reddening vector, while the continuous and dashed lines are the
empirical ZAMS from Schmidt-Kaler (1982).
Cluster probable members are indicated with filled symbols, while non members with
open symbols. The position of HD~92207 is indicated with a starred symbol. See the text for
details}
\end{figure}

\begin{figure}
\centerline{\psfig{file=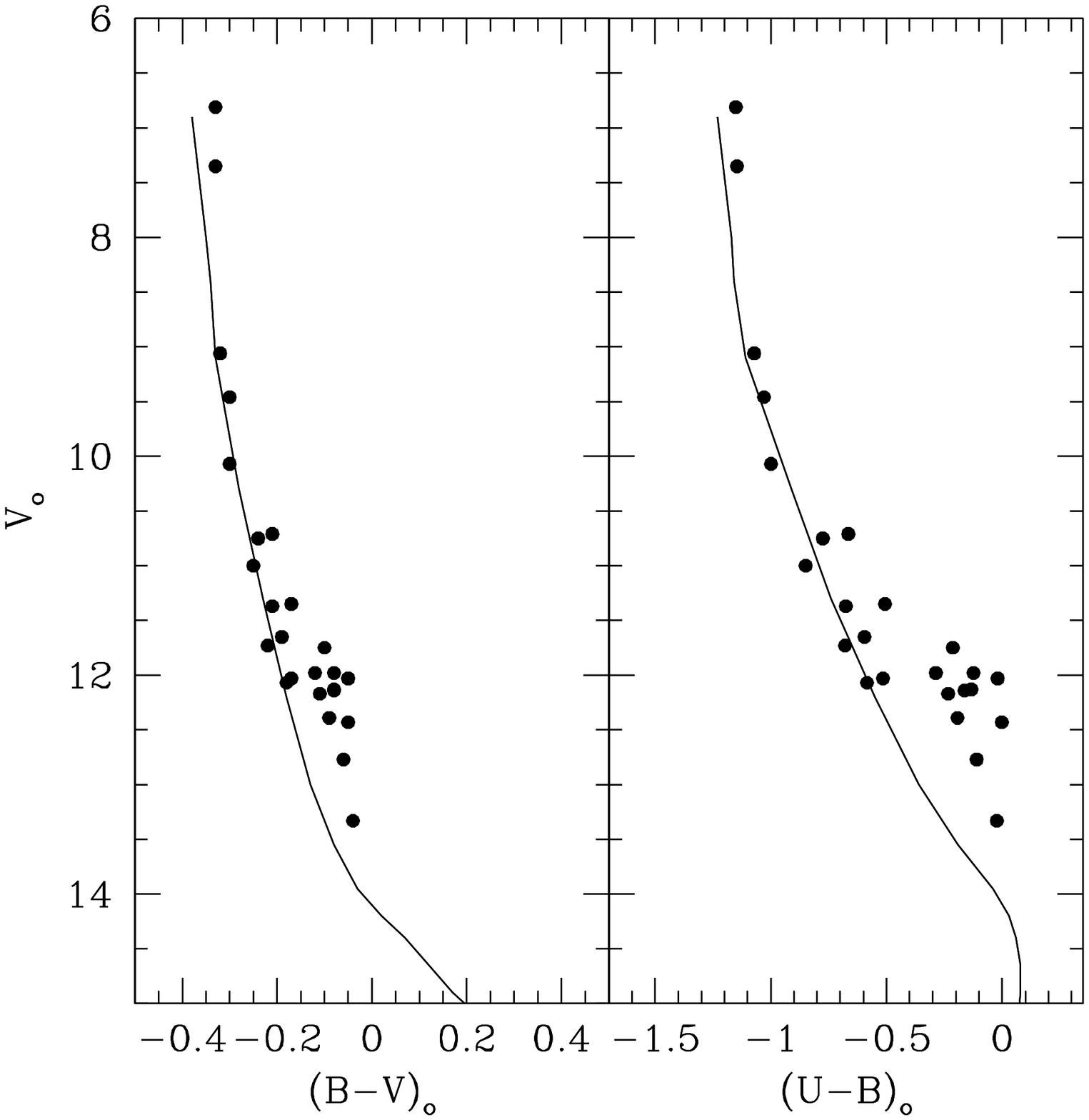,width=10cm,height=10cm}}
\caption{The reddening corrected CMDs of NGC~3324. Only probable
members are plotted,  together with the Schmidt-Kaler sequence. See text for details.}
\end{figure}

\subsection{Reddening}
The position of the studied stars in the two color diagram is shown in
Fig.~3. In the same plot we show an un-reddened sequence (solid line)
from Schmidt-Kaler (1982), and the same sequence shifted by E$(B-V)$~=~0.48
and E$(U-B)$~=~0.34 (dashed line). Most of the stars actually lie along
this second ZAMS, sharing the same reddening of  E$(B-V)$~=~0.48$\pm$0.03.
All these stars are considered probable members. 
They amount at 25 stars (filled circles)
up  to $V$~=~15, and have been collected in Table~3.
The remaining stars are probably non-members.\\
Our estimate for  the reddening is consistent with that obtained
by Clari\'a (1977). HD~92207 (starred symbol) lies somewhat outside the cluster mean line, and this raises some doubts on its membership to NGC~3324.

\begin{figure}
\centerline{\psfig{file=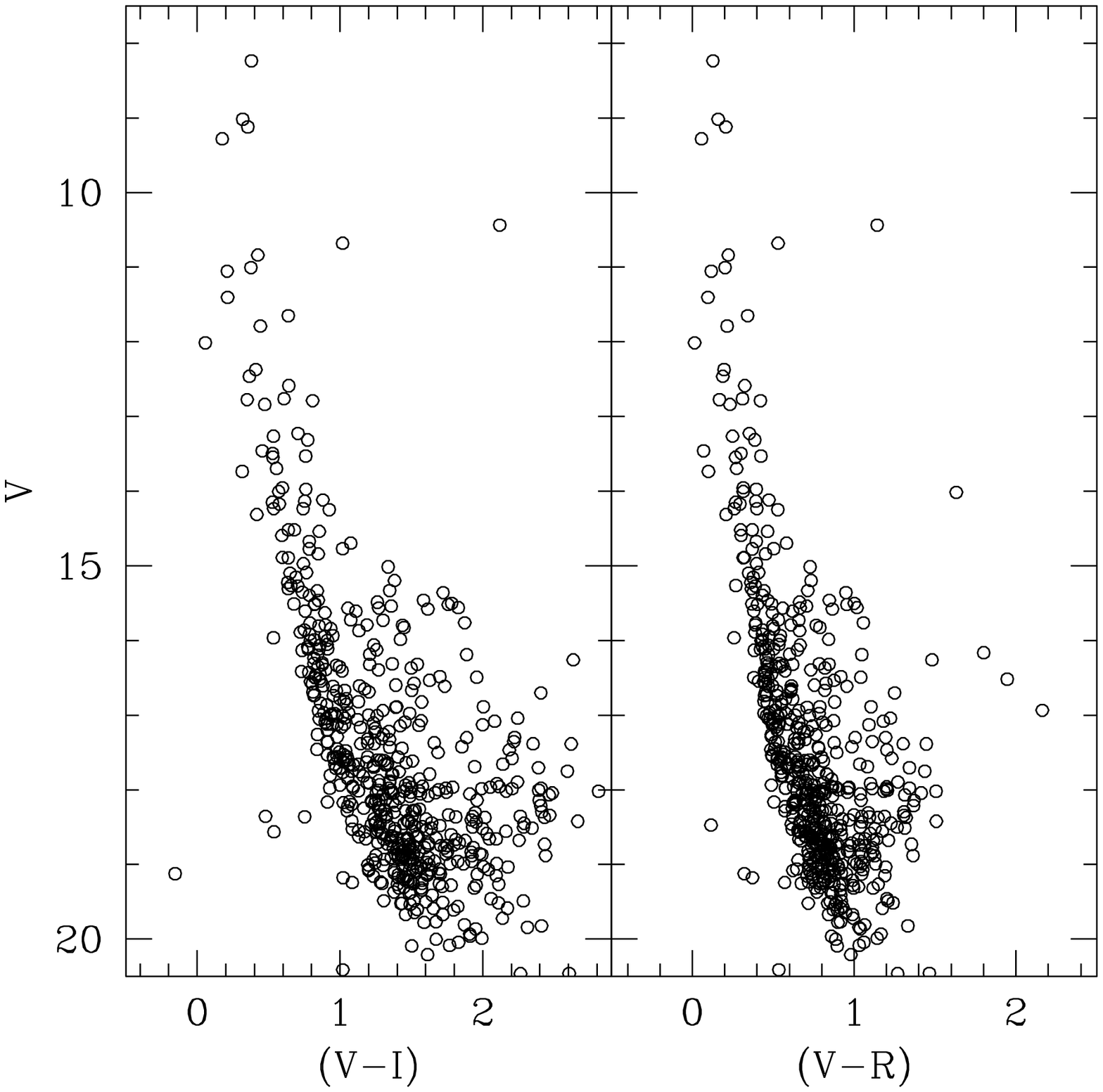,width=10cm,height=10cm}}
\caption{The CMDs of NGC~3324 for all the detected stars.}
\end{figure}

\subsection{Age and distance}
The Color Magnitude Diagrams (CMDs) of NGC~3324 in different colors 
are shown in Figs.~4 and 5. Fig.~4 shows the reddening
corrected  CMD in the planes
$V_o -(B-V)_o$ (left panel) and $V_o-(U-B)_o$ (right panel), where only the
probable cluster members have been plotted. super-imposed is the empirical
ZAMS from Schmidt-Kaler (1982) shifted by E$(B-V)$~=~0.48
and E$(U-B)$~=~0.34, respectively. The vertical shift provides
an estimate of the apparent distance modulus $(m-M)$ = 13.9$\pm$0.03,
where the uncertainty has been computed allowing the reddening
to lie in the range derived in the previous section. 
Therefore the corrected distance modulus $(m-M)_o$ turns out to be 
12.41$\pm$0.03,
in agreement with Clari\'a's (1977) estimate. 
Accordingly, the distance of NGC~3324 to the Sun is 3.0$\pm$0.1 kpc.\\
On the other hand, Fig.~5 shows the CMD in the planes
$V-(V-I)$ and $V-(V-R)$. Note the extension of the MS down to V~=~20,
and the widening of the MS at increasing magnitudes. Since the width of the MS
below $V$~=~14 is much larger than  the mean photometric errors,
it is quite probable that many pre-MS stars are actually present in this cluster (see below).\\

Estimating the precise age of this cluster is a challenging task.
Looking again at the CMDs, apparently there are no bright stars
which are significantly redder than the MS. This means
that all the stars we see are still in the H-burning  phase.\
The exact age of the cluster strongly depends on the membership
of HD~92207. If this star belongs to the cluster, it must be
an evolved star due to its red color. In this case  the cluster
would be as old as 2-3 Myr, as Clari\'a (1977) pointed out.
On the other hand, if HD~92207 does not belong to the cluster,
the 2-3 Myr estimate has to be considered only as an upper limit
for the age of NGC~3324.

\begin{figure*}
\centerline{\psfig{file=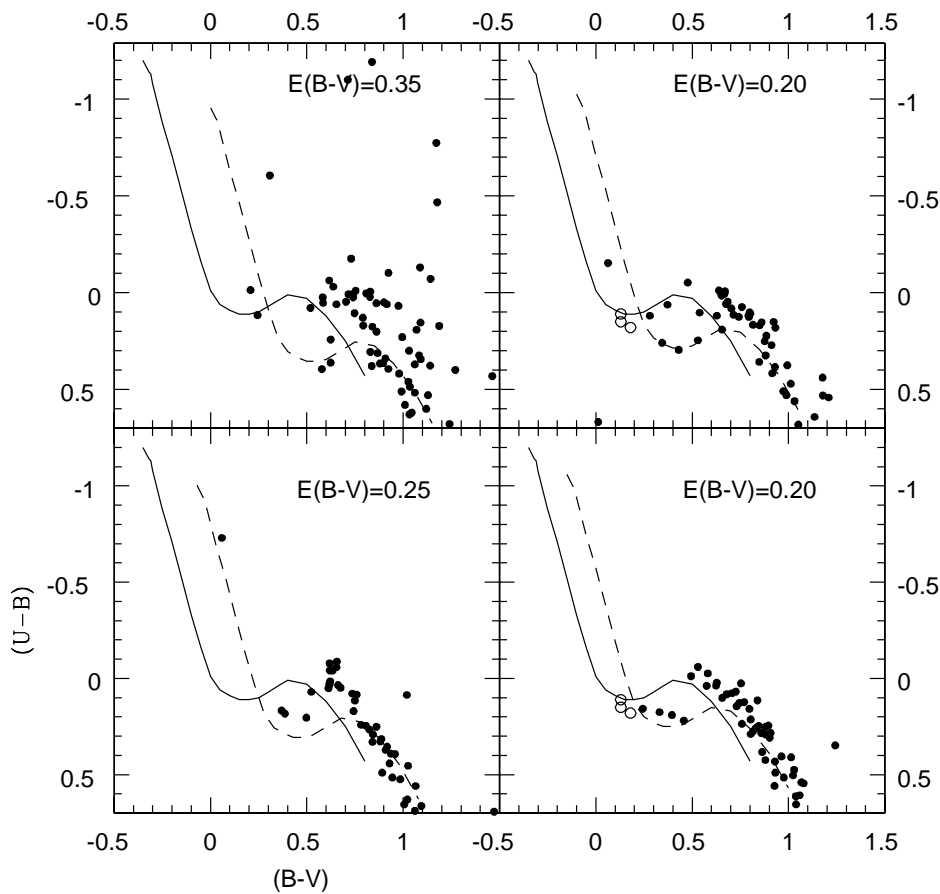,width=14cm,height=14cm}}
\caption{Two colors diagrams for all the stars
in the region of Loden~165, grouped according to the targeted
fields (see Table~1). In the upper panels two different sequences are
clearly visible. Open symbols indicate the stars studied by Lod\'en (1973).}
\end{figure*}

\subsection{Looking for pre-MS candidates}
Looking carefully at Fig.~4, it seems that the clump of stars with
$V \le 14.0$ lies away from the ZAMS towards the red. This may be an
indication that these stars are still in the pre MS phase ad are still
approaching the ZAMS. In addition the appearance of the CMDs in Fig~5
points towards the same direction. The MS gets wider at increasing
magnitude, and its width is significantly larger  than 
expected from the photometric errors. Moreover
since we are covering a rather small region of the cluster,
the contamination of field stars should not be very significant.\\
Much firmer conclusions can be obtained by measuring
the $H_{\alpha}$ emission from these candidates or by obtaining infrared
photometry (Hillebrandt et al 1998).

\subsection{Absolute proper motion and the membership of HD~92207}
The Hipparcos Catalogue (ESA 1997) provides absolute proper motions for
120.000 stars, covering the whole sky. It is supplemented to
fainter magnitudes by the Tycho 2 Catalogue (H\o g et al.\ 2000).
The proper motion of HD~92207 was measured by Hipparcos and we could identify
4 other potential members of NGC~3324 in Tycho~2. All stars are listed
in Table~3, and are considered probable members on the basis
of the position in the CMD - they lie along the MS- and because they 
have the same reddening.
From the Tycho~2 members, the mean cluster 
motion turns out to be
$\mu_{\alpha*}~=~-10.50\pm1.19$ and  $\mu_{\delta}~=~3.79 +- 1.16$ $mas/yr$.
There is an agreement with the proper motion of HD~92207 in $\mu_{\delta}$, but 
a significant discrepancy in $\mu_{\alpha}$. 
Comparing the proper motions, we derive a membership probability 
of $5.5\%$ for HD~92207
(see Baumgardt et al 2000 for the method to assign probability)
, which argues against the  membership of this star in the cluster. 
However, the mean cluster motion is not very well defined,
since it is derived from four stars only.
The proper
motion alone does therefore not rule out the membership of HD~92207.\\

\begin{table}
\caption{Proper motions (in $mas/yr$) of possible members of NGC~3324. 
Data is taken from Tycho 2, except for
HD~92207, which has been taken from Hipparcos.}
\begin{tabular}{l@{\hspace{2pt}}rccc}
\hline
\hline
\multicolumn{1}{c}{ID} &
\multicolumn{2}{c}{Star} &
\multicolumn{1}{c}{$\mu_{\alpha*}$}  &
\multicolumn{1}{c}{$\mu_{\delta}$} \\
\hline
 1 & GSC 8613 & 1825 & -10.4 $\pm$ 4.2  &  6.8$\pm$ 4.0 \\
 2 & GSC 8613 & 780  & -8.0  $\pm$ 1.5  &  5.4$\pm$ 1.5  \\
 6 & GSC 8613 & 121  & -16.5 $\pm$ 3.6  &  1.7$\pm$ 3.4  \\
 7 & GSC 8613 & 947  & -15.6 $\pm$ 2.8  & -1.1 $\pm$ 1.6  \\
   & HD~92207 &      & -7.46 $\pm$ 0.53 & 3.22 $\pm$ 0.44   \\
\hline\hline
\end{tabular}
\end{table}

\noindent
HD~92207 seems also to have a higher reddening (E$(B-V)$~=~0.52) than the mean NGC~3324
value (0.48) , and according to Clari\'a (1977) a lower distance (1.9~kpc).\\

Summarizing, we conclude that HD~92207 is unlikely to be a genuine member of NGC~3324. 
It might however be spatially close to this cluster, since the HII gas in this area
surrounds the star and seems to connect it with the cluster center.
But this may just be a projection effect.

\section{The star cluster Loden 165}
Nothing exists in the literature about this cluster,
designed as C1035-584 by the IAU.
It was suggested to be a possible physical group
by Lod\'en(1973) who 
measured three stars (HDE~303088, HDE~303089 
and HDE~303090), and found that they are of A0-A2 spectral type.
Lod\'en  suggests that this triplet is 1 kpc distant from the Sun
and suffers from a low extinction, which he found to be $E(B-V)$~=~0.11.\\

\begin{table}
\caption{Basic parameters of Loden~165 stars.}
\begin{tabular}{ccccc}
\hline
\hline
\multicolumn{1}{c}{Name} &
\multicolumn{1}{c}{$V$}  &
\multicolumn{1}{c}{$(B-V)$}  &
\multicolumn{1}{c}{$(U-B)$} &
\multicolumn{1}{c}{$Sp. Type$} \\
\hline
HD~303088      & 11.20 & 0.13  & 0.15 & A2  \\
HD~303089      & 11.39 & 0.13  & 0.11 & A0  \\
HD~303090      & 11.35 & 0.18  & 0.18 & A2  \\
\hline\hline
\end{tabular}
\end{table}

Since this triplet appears rather compact, the stars might in principle
constitute a poor open cluster or an association.
The coordinates of the triplet are given in Table~1, from which
it can be seen that Loden~165 lies about $4^{\prime}$ west
of our second field. This can also be seen in Fig.~1, where Lod\'en's stars 
are enclosed in a small circle, whereas the field covered by us
is identified by the lower right square.

The area covered by our observations does not include Loden~165 stars. However
this field is very interesting due to its vicinity to NGC~3324.
According to the estimate of its tidal radius ($12^{\prime}$),
the field covered by us could simply be a portion
of the halo of NGC~3324.\\
Alternatively, we could be simply looking at a region
of Loden~165, whose nature of cluster has never been clarified.

\subsection{The two colors diagram}
The two color diagram for the stars in the field of Loden
is shown in Fig.~6, where we have divided the regions according to the 
observed fields which, combined together, cover the region shown in Fig.~1. 
It is clear that the region is dominated by
differential reddening. In particular in the upper
frames the stars are rather dispersed, and seem to show several
parallel sequences,
characterized by a different mean reddening.
The larger value of the reddening in this region with respect to the other
fields,
is marginally compatible with the reddening of HD~92207 and NGC~3324, which lie
very close.\\
On the opposite side the reddening is lower and close
to the one reported by Lod\'en for his triplet.
However, on the average, the mean reddening of this region is much lower 
than in the case of   NGC~3324 (cfr. Section~3.3), being between $0.20-0.35$.
The mean reddening in this region,
averaging over all four fields, turns out  to be $E(B-V)=0.25\pm0.08$.\\
The three stars of Loden~165 are shown in the right panels 
of Fig.~6 with open symbols.
They seem to lie somewhat apart from the ZAMS in both
panels, supporting the conclusion that they are a triplet of stars lying
closer to the Sun and suffering from a smaller reddening. 

\begin{figure}
\centerline{\psfig{file=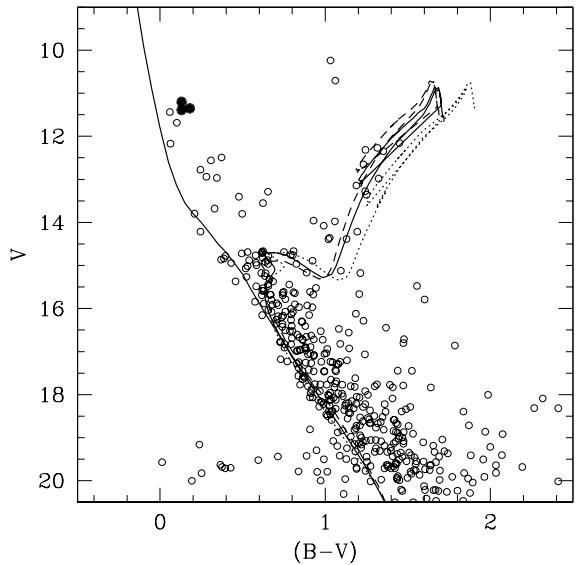,width=9cm,height=9cm}}
\caption{CMD in the region of Loden~165. Superimposed
are  three solar    metallicity isochrones
for the ages of 2 (dotted line), 3 (solid line)
and 4 (dashed line) Gyrs (Girardi et al 2000).
The three filled  circles indicate  Lod\'en's triplet.}
\end{figure}

\subsection{The CMDs}
The global CMD of Loden~165 is shown
in Fig.~7. The general appearance is very different
from the CMD of NGC~3324 (cfr. Figs.~4 and 5). There is a clear
MS extending for about 5 magnitudes from $V$~=~20 to $V$~=~15. At 
brighter magnitudes it sharply drops, showing a clear
Turn Off at $V$~=~14.5, $(B-V)$~=~0.25. The population
of bright MS stars is very poor, while there is 
clear evidence of a sub-giant and red giant branch
population. This leads to the suggestion that
the cluster is rather old, at least much older
than NGC~3324.\\
To have an idea of the cluster age and distance
we superimposed three solar metallicity isochrones
for the ages of 2 (dotted line), 3 (solid line) and 
4 (dashed line) Gyrs from the Padova models (Girardi et al. 2000).
It turns out that the best match is obtained with the 3 Gyrs isochrone
by assuming the mean reddening
derived above.
As a consequence, the apparent distance modulus turns
out to be (m-M)~=~12.1. The corrected distance
modulus turns out to be $(m-M)_o~=~11.42$, 
 which puts this object
1.9 kpc from the Sun.
In the same plot we have included  also the Lod\'en triplet
using filled circles. 
The most plausible conclusion is that they seem to be simply
field stars.

\section{Conclusions}
We reported on $UBVRI$ photometry of two stellar fields in the
open cluster NGC~3324. Our results basically confirm
previous photo-electric
investigations. We find 25 probable members and suggest that
the cluster is younger than 3 {\it Myr}, suffers from a mean extinction
of E$(B-V)$~=~0.48$\pm$0.03, and lies 3.0$\pm$0.1 kpc from the Sun. 
By extending
the member list up to $V$~=~14.5, we suggest the possibility
that some stars are actually pre-MS candidates, an hypothesis
which deserves  
further investigation to be confirmed. Pre-MS
objects in fact would  provide an additional constraint
on the cluster age. The A super-giant HD~92207 is probably
not a member of NGC~3324, but might be spatially
close to the cluster.\\

Loden~165, an object never studied so far, is an old
star cluster with an age of approximately 3 Gyrs.   
The estimates of the age, reddening and distance are very different 
from those of NGC~3324, suggesting that Loden~165 has 
presumably nothing to do with the Carina open cluster system.

\begin{acknowledgements}
GC acknowledges kind hospitality from ESO,
and expresses his gratitude to M. Romaniello, 
L. Pasquini and A. Brown for stimulating discussions.
This study made use of Simbad and WEBDA. 
\end{acknowledgements}

{}

\end{document}